# Softened $sp^2$-$sp^3$ bonding network leads to strong anharmonicity and weak hydrodynamics in graphene+

Linfeng Yu,[1] Ailing Chen,[1] Xiaoxia Wang,[2] Huimin Wang,[2] Zhenzhen Qin,[3] and Guangzhao Qin[1,*]

[1]*State Key Laboratory of Advanced Design and Manufacturing for Vehicle Body, College of Mechanical and Vehicle Engineering, Hunan University, Changsha 410082, People's Republic of China*
[2]*Hunan Key Laboratory for Micro-Nano Energy Materials & Device and School of Physics and Optoelectronics, Xiangtan University, Xiangtan 411105, Hunan, China*
[3]*School of Physics and Microelectronics, Zhengzhou University, Zhengzhou 450001, China*

Graphene+, a novel carbon monolayer with $sp^2$-$sp^3$ hybridization, was recently reported to exhibit unprecedented out-of-plane half-auxetic behavior and graphenelike Dirac properties [Yu *et al.*, Cell Rep. Phys. Sci. 3, 100790 (2022)]. Herein, based on comprehensive state-of-the-art first-principles calculations, we reveal the significant effect of softened $sp^2$-$sp^3$ bonding on the lattice thermal transport in graphene+. At room temperature, the thermal conductivity ($\kappa$) of graphene+ is obtained as ∼170 W/mK, which is much lower than that of graphene (∼3170 W/mK). It is found that the softened $sp^2$-$sp^3$ bonding significantly suppresses the vibrations of acoustic phonons in graphene+, which leads to strong anharmonicity and weak phonon hydrodynamics. Thus, the large reduction in $\kappa$ stems from the softened $sp^2$-$sp^3$ bonding network. Our study provides fundamental physical insights into the thermal transport properties of graphene+, which would provide prospective guidance for the promising application in the field of thermal management.

## I. INTRODUCTION

Recent decades have witnessed the vigorous development of carbon materials, and many excellent carbon materials have been reported [1], including diamond [2], T-carbon [3], and numerous two-dimensional (2D) structures centered on graphene [4–8], which exhibit excellent properties [9–11]. Typically, graphene is widely used as a research benchmark in monolayer carbon materials due to its outstanding properties, such as electronic Dirac properties [12], high breaking strength [13], and ultrahigh thermal conductivity ($\kappa$) [14]. Extensive efforts have been devoted to searching for promising 2D carbon candidates beyond graphene. For instance, Zhang *et al.* found that pure pentagonal rings can form a novel 2D carbon allotrope, namely penta-graphene [15], which serves as an excellent design platform for 2D penta-materials [16–19]. Moreover, a series of 2D carbons with outstanding properties have been reported, such as T-graphene [20], Twin-graphene [21], SW-graphene [6], and biphenylene [22,23]. Abundant configurations enable carbon materials to form a promising candidate library for micro-nano electronic devices. In practical applications, the thermal transport property is of great significance to the working lifetime and performance of micro-nano electronic devices for its key role in efficient heat dissipation, which demands fundamental study. However, despite the boom in carbon materials, insights into thermal transport properties remain challenging because of the high cost of experimental equipment and computational resources.

With the rich and diverse bonding configurations in carbon materials, an in-depth understanding of thermal transport properties in 2D carbon materials is becoming a central issue. For 2D carbon materials, the strong planar $sp^2$ hybrid C-C bond properties in graphene led to its ultrahigh $\kappa$ of 3000–5000 W/mK [14,24,25]. Choudhry *et al.* reported that the reduced $\kappa$ of graphene allotropes is due to phonon branches folding induced by the phononic crystal, independent of bond strength [26]. In addition, T-graphene, D-graphene, and biphenylene have been reported to exhibit distinctive thermal transport behaviors due to a diverse $sp^2$ bonding arrangement, with $\kappa$ being 800, 600, and 166 (254) W/mK, respectively [22,26]. However, the $sp^3$ hybridization effect is not considered in these studies. It has been reported that $sp^3$ hybridization induces a strongly buckling structure in penta-graphene, which contributes to the large scattering phase space and low $\kappa$ (645 W/mK) due to the broken *out-of-plane* symmetry [27]. Such a phenomenon also generally occurs in other 2D buckling compounds, such as silicene [25], g-$B_3N_5$ [28], and others [29,30]. Interestingly, it is not found in our study, despite the similar buckling structures and pentagonal rings. Those diverse thermal transport behaviors largely originate from the different C-C bond morphology. With the exception of penta-graphene, with its $sp^2$-$sp^3$ hybridization, previously reported studies involving thermal transport properties have mainly focused on carbon monolayers with $sp^2$ hybridization bonding networks. It remains poorly understood how the complex $sp^2$-$sp^3$ (or $sp^3$) hybrid bonding affects the thermal transport properties.

---

[*]Author to whom all correspondence should be addressed: gzqin@hnu.edu.cn

Recently, an $sp^2$-$sp^3$ hybrid carbon monolayer graphene-plus (graphene+) has been reported to possess an unprecedented novel *out-of-plane* half-auxetic behavior in our study [31]. The novel auxetic behavior caused by the $sp^3$ hybridization-induced geometric mode switching beyond graphene is expected to bring more interesting applications and prospects for graphene+, such as shock absorption and energy absorption [32–34]. In addition, graphene+ also exhibits outstanding electronic Dirac properties, which typically exhibit the quantum anomalous Hall effect and high carrier mobility [35–38]. It is hoped that, as with theoretical and experimental researchers who cooperated to predict and synthesize novel materials, future studies will synthesize graphene+. For instance, monolayer amorphous carbon [39], biphenylene [23], and two-dimensional fullerene frameworks [40] have been experimentally prepared. In addition, T-carbon was first predicted in 2011 [3], and it was experimentally synthesized in 2017 [41]. Note that graphene+ is energetically preferable over the already experimentally synthesized T-carbon [31]. These successful examples can provide valuable clues for the synthesis of graphene+. However, the unknown thermal transport properties and the fundamental understanding of its mechanism limit its further development and possible applications. The hybrid $sp^2$-$sp^3$ bonding in graphene+ is expected to shed light on the fundamental understanding of the bonding effect on thermal transport.

In this paper, based on state-of-the-art first-principles calculations, we explored comprehensively the effect of complex hybrid bonding on thermal transport properties by taking graphene+ as a study case, with graphene as a benchmark comparison. One order of magnitude lower $\kappa$ than graphene and exceptional thermal transport properties are found in graphene+. To reveal the origin, we conduct mode-level phonon property images and discuss the effects of lattice dynamics, four-phonon scattering, size effects, and phonon hydrodynamics, which highlight the key role of quantified $sp^2$-$sp^3$ bonding in thermal transport. The mechanism as revealed in this study is expected to provide an answer to the open question of how atom bonding affects thermal transport, which would promote the applications of graphene+ and other carbon materials in electronics with high-performance thermal management.

## II. COMPUTATIONAL METHODOLOGY

All first-principles calculations are performed with the aid of the Vienna *ab initio* simulation package (VASP) [42,43] based on density functional theory (DFT). The Perdew-Burke-Ernzerhof (PBE) functional [44] is used to describe the exchange and correlation interactions between electrons. The kinetic energy cutoff of 1000 eV, 2.5 times the recommended value for carbon in the potential file, is selected to expand the wave functions in a plane-wave basis set. A Monkhorst-Pack [45] $q$-mesh of $9 \times 9 \times 1$ ($15 \times 15 \times 1$) is used to optimize the crystal structure for graphene+ (graphene) with the accuracy of $10^{-6}$ and $10^{-5}$ eV for energy and Hellmann-Feynman force convergence, respectively. The $2 \times 2 \times 1$ ($5 \times 5 \times 1$) supercell with $2 \times 2 \times 1$ $q$-mesh is used to calculate second-order and third-order force constants for graphene+ (graphene). Based on harmonic and anharmonic force constants, the $\kappa$ is obtained with the Sheng-BTE [46] code,

$$\kappa = \kappa_{\alpha\alpha} = \frac{1}{V}\sum_{\lambda} C_\lambda v_{\lambda\alpha}^2 \tau_{\lambda\alpha}, \quad (1)$$

where $\tau_{\lambda\alpha}, C_\lambda, v_{\lambda\alpha}$, and $V$ are the relaxation time, the specific-heat capacity, the group velocity for $\lambda$ mode phonon along the $\alpha$ direction, and the crystal volume, respectively.

TABLE I. Comparison of hydrodynamic effects in different hybridized two-dimensional carbon structures.

|  | $\kappa_{\text{ITE}}$ | $\kappa_{\text{RTA}}$ | H = $\kappa_{\text{ITE}}/\kappa_{\text{RTA}}$ | Bond network |
|---|---|---|---|---|
| Graphene+ | 170 | 110 | 1.56 | $sp^2$-$sp^3$ |
| Penta-graphene | 437 | 272 | 1.61 | $sp^2$-$sp^3$ |
| Biphenylene | 267 (x) | 142 (x) | 1.88 (x) | $sp^2$ |
|  | 421 (y) | 207 (y) | 2.03 (y) |  |
| T-graphene | 645 | 243 | 2.65 | $sp^2$ |
| Graphene | 3170 | 440 | 7.20 | $sp^2$ |

## III. RESULTS

### A. Effects of $sp^2$-$sp^3$ hybridization

The top and side views of the crystal geometry for graphene and graphene+ are comparably shown in Figs. 1(a) and 1(b). As a typical $sp^2$ hybrid carbon monolayer, each carbon atom in graphene forms strong C-C bonds with three other carbon atoms. The difference between graphene+ and graphene is that the five-membered rings in graphene+ are distributed along the diagonal of the square to form a carbon network with $sp^2$-$sp^3$ hybridization. The two $sp^3$ carbon atoms are in the center of the diagonal line, and the $sp^2$ carbon atoms form a plus "+" shaped distribution around the $sp^3$ carbon atoms, which is why it is called graphene-plus (graphene+). Since the bonding in graphene+ deviates from the ideal $sp^2$ hybridization in graphene, out-of-plane half-auxetic behavior emerges in graphene+ despite exhibiting Dirac properties [31]. In particular, the thermal transport properties of graphene+ are also expected to exhibit distinct extraordinary properties as a complement and candidate beyond graphene in the field of thermal management due to the emergence of $sp^3$ hybridization.

To reveal the effect of $sp^2$-$sp^3$ hybridization on thermal transport behavior, the $\kappa$ of several typical 2D carbon allotropes are calculated at 300 K, including penta-graphene [$sp^2$-$sp^3$, Fig. 1(c)], biphenylene [$sp^2$, Fig. 1(d)], and T-graphene [$sp^2$, Fig. 1(e)]. As comparably shown in Table I and Fig. 1(f), it is found that graphene+ exhibited the lowest $\kappa$ of 170 W/mK, while the $\kappa$ of graphene, T-graphene, biphenylene, and penta-graphene are 3170, 645, 421 (267), and 437 W/mK, respectively. Lower $\kappa$ means stronger anharmonicity in graphene+. Furthermore, as an important feature in the thermal transport properties for graphene allotropes, phonon hydrodynamics plays a vital role in improving the $\kappa$ [31]. And it can be evaluated by comparing the $\kappa$ from iterative (ITE) and relaxation time approximation (RTA) methods [25,47]. The $\kappa$ ratio H = $\kappa_{\text{ITE}}/\kappa_{\text{RTA}}$ of ITE to RTA methods can be used to measure the strength of phonon hydrodynamics. The higher $H$ means a stronger hydrodynamics effect. As shown in Fig. 1(f), smaller $H$ values are found in 2D $sp^2$-$sp^3$ carbon allotropes compared with the $sp^2$ hybridized systems, i.e., $H_{gra+} = 1.56 < H_{penta} = 1.61 < H_{bip} = 1.88$ or $2.03 < H_{T-gra} = 2.65 < H_{gra} = 7.20$, indicating that the introduction of $sp^3$ hybridization significantly depresses phonon hydrodynamics. Intriguingly, stronger anharmonicity and weaker hydrodynamic effects are found in graphene+ due to $sp^3$ hybridization. Such an interesting phenomenon motivates us to further conduct a deep comprehensive study on the thermal transport behavior of graphene+ as presented in the following.

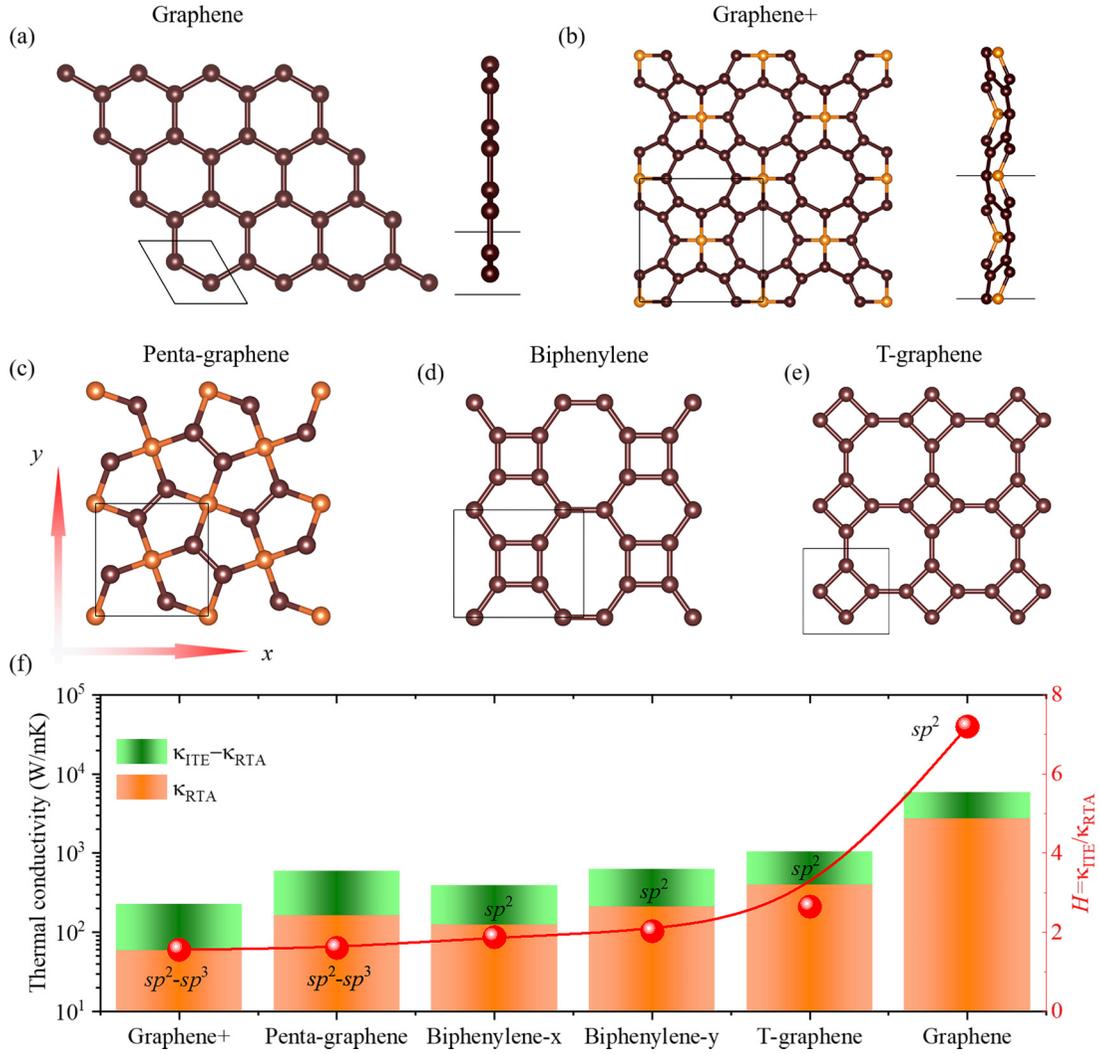

FIG. 1. (a) Top and side views of graphene. (b) Top and side views of graphene+. The top views of (c) penta-graphene, (d) biphenylene, and (e) T-graphene. (f) The effect of $sp^3$ hybridization on thermal conductivity. The "ITE" and "RTA" denote iterative and relaxation time approximation methods, respectively.

### B. Interatomic interactions

The thermal transport is fundamentally determined by the bonding strength and interaction range, which can be evaluated by the normalized force constant trace (FCT). Based on the finite displacement difference method in real space, the interatomic force constant can obtain as the second derivative of the energy $E$ by [48,49]

$$\frac{\partial^2 E}{\partial R_i \partial R_j} = \begin{bmatrix} \frac{\partial^2 E}{\partial R_x \partial R_x} & \frac{\partial^2 E}{\partial R_x \partial R_y} & \frac{\partial^2 E}{\partial R_x \partial R_z} \\ \frac{\partial^2 E}{\partial R_y \partial R_x} & \frac{\partial^2 E}{\partial R_y \partial R_y} & \frac{\partial^2 E}{\partial R_y \partial R_z} \\ \frac{\partial^2 E}{\partial R_z \partial R_x} & \frac{\partial^2 E}{\partial R_z \partial R_y} & \frac{\partial^2 E}{\partial R_z \partial R_z} \end{bmatrix}, \quad (2)$$

where $R_\alpha$ represents the position of the atom along the $\alpha$ direction. The normalized FCTs are calculated to assess the relative strength of the bonding between different atoms with different spacing [48,49], thereby further determining their interaction range [48]:

$$\text{FCT} = \frac{\partial^2 E}{\partial R_x \partial R_x} + \frac{\partial^2 E}{\partial R_y \partial R_y} + \frac{\partial^2 E}{\partial R_z \partial R_z}, \quad (3)$$

$$\text{normalized FCT} = \frac{\frac{\partial^2 E}{\partial R_{0,x} \partial R_{n,x}} + \frac{\partial^2 E}{\partial R_{0,y} \partial R_{n,y}} + \frac{\partial^2 E}{\partial R_{0,z} \partial R_{n,z}}}{\frac{\partial^2 E}{\partial R_{0,x} \partial R_{0,x}} + \frac{\partial^2 E}{\partial R_{0,y} \partial R_{0,y}} + \frac{\partial^2 E}{\partial R_{0,z} \partial R_{0,z}}}, \quad (4)$$

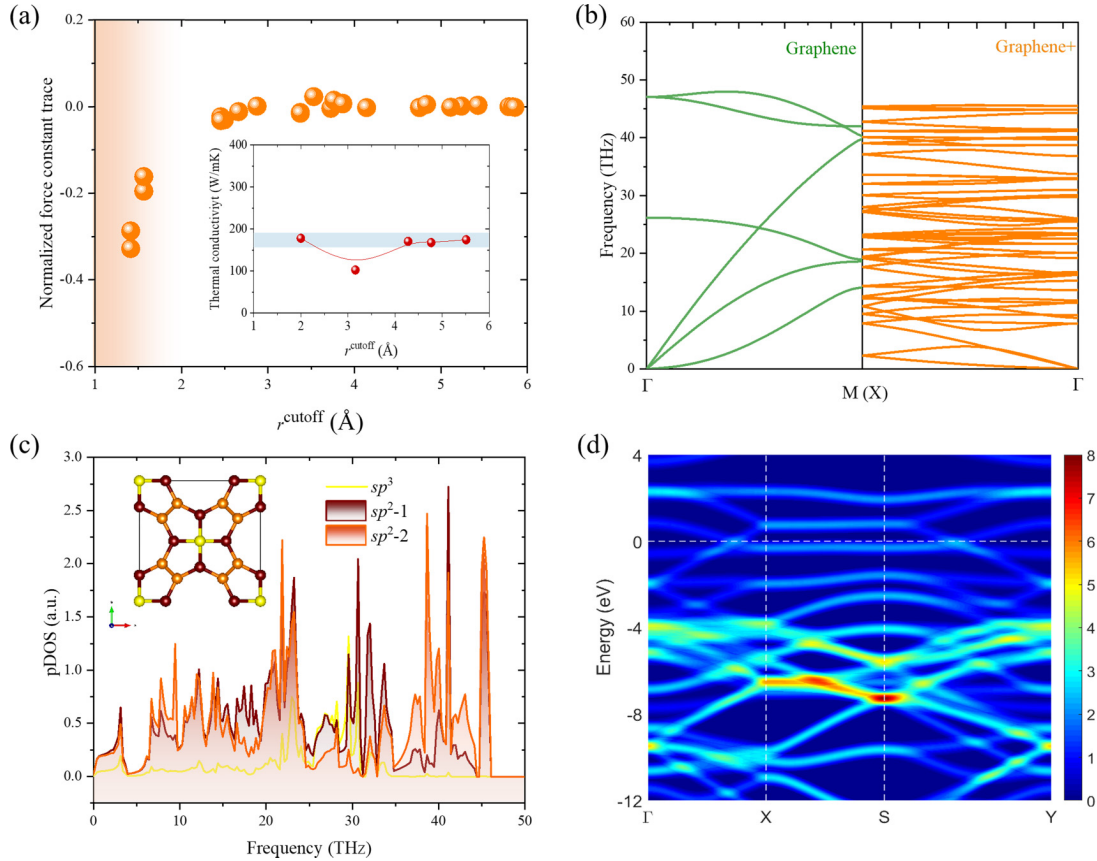

FIG. 2. (a) Normalized force constant trace, and the inset is that thermal conductivity at 300 K as a function of cutoff radius ($r^{cutoff}$ in Å) for graphene+. The light blue shading marks a 10% error area relative to the $\kappa$ at $r^{cutoff}$ = 5.4 Å. (b) The comparison of phonon dispersion for graphene and graphene+. (c) The partial density of states (pDOS) of graphene+. The inset marks the atoms in $sp^3$ and $sp^2$ hybridization with different colors. (d) Inhomogeneous electron distribution in the energy bands of graphene+.

where $\frac{\partial^2 E}{\partial R_{i,x} \partial R_{j,\alpha}}$ denotes the harmonic force constant of the interaction between the $i$th atom and the $j$th atom along the $x$ direction. When $i = j = 0$, $\frac{\partial^2 E}{\partial R_{i,x} \partial R_{j,\alpha}}$ represents the self-interaction force constant along the $\alpha$ direction. Correspondingly, when $i = 0$, $j = n$, it represents the force constant of the atom of the $n$th nearest neighbor relative to the initial (labeled "0") atom. The cutoff radius ($r^{cutoff}$)-dependent normalized FCT behavior enables the evaluation of long-range interactions between atoms, which successfully explained the long-range effect of the resonant bonding in the rocksalt structure, such as bulk Pb$X$ ($X$ = S, Se, and Te) and monolayer black phosphorene [48,50–53].

According to Eqs. (2)–(4), the result of the normalized FCT for graphene+ is plotted in Fig. 2(a). When the $r^{cutoff}$ exceeds 2.5 Å, the normalized FCT values display obvious convergence. As the $r^{cutoff}$ increases, the normalized FCTs do not change significantly, especially when $r^{cutoff}$ > 4 Å. Such behavior implies weak long-range interactions in graphene+. The reason may lie in the fact that the introduction of $sp^3$ hybridization in graphene+ destroys the original ring-shaped large $\pi$ bond to form a brand new bonding morphology, while the $\pi$ bond induces a long-range resonance in graphene [34]. In addition, the $r^{cutoff}$-dependent $\kappa$ behavior also suggests weak long-range interactions, as shown in the inset of

Fig. 2(a). The slight error in the $\kappa$ oscillations may result from the initial random guess of the charge density in DFT calculations. Therefore, 5.4 Å is selected as the effective $r^{cutoff}$ to obtain a satisfactory and strictly convergent $\kappa$, which is large enough to include the weak long-range interactions and the possible anharmonic effects.

### C. Competitive mechanism of lattice dynamics

The lattice dynamics and related thermal transport properties for graphene and graphene+ can be evaluated based on the phonon dispersion, which is plotted in Fig. 2(b). The total number of phonon branches "$N_{ph}$" depends on the number of atoms in the unit cell "$n$" by the standard relation $N_{ph} = 3n$ ($n = 2$ for graphene, and $n = 18$ for graphene+). Hence, the results of $N_{ph}$ = 6 and 54 are found in graphene and graphene+, respectively, indicating more abundant phonon branches in graphene+ (see Fig. S1 in the Supplemental Material [54]). Besides, almost the same phonon vibrational frequency range (0–50 THz) is manifested in the phonon dispersion for graphene and graphene+. However, due to the unique $sp^2$-$sp^3$ network, more complex phonon dispersion appears in graphene+, raising a competing mechanism, namely the phonon beam and softened phonons. The concentrated phonon bands appear flattened in the phonon dispersion of graphene+, leading to the phonon beam effect [55]. With the phonon beam effect, the small gap between the phonon bands makes it difficult for three phonons with close frequencies to be scattered (see Note S1 and Fig. S3 in the Supplemental Material [54]), whether absorbed or emitted [29]. As a result, it can effectively reduce the phonon scattering phase space [Fig. 4(c)] and improve $\kappa$. Simultaneously, the three acoustic phonon branches at low frequencies are depressed below 10 THz due to the substantial filling of the phonon band. The maximum phonon frequencies ($w_A$) of the acoustic phonon branch at the high symmetry point $M$ or $X$ are 39.8 and 7.9 THz for graphene+ and graphene, respectively. Based on the equation of $\theta_D = hw_A/k_B$ ($h$ and $k_B$ are Planck's constant and Boltzmann's constant, respectively), lower Debye temperature $\theta_D$ is obtained in graphene+ (378 K) compared to graphene (1910 K). It is well known that frequency-reduced acoustic phonon modes are more easily scattered in thermal vibrations due to lower Debye temperature. Therefore, the competing mechanism of phonon beam and softened phonons in graphene+ is expected to lead to exceptional thermal transport behavior, calling for comprehensive investigations.

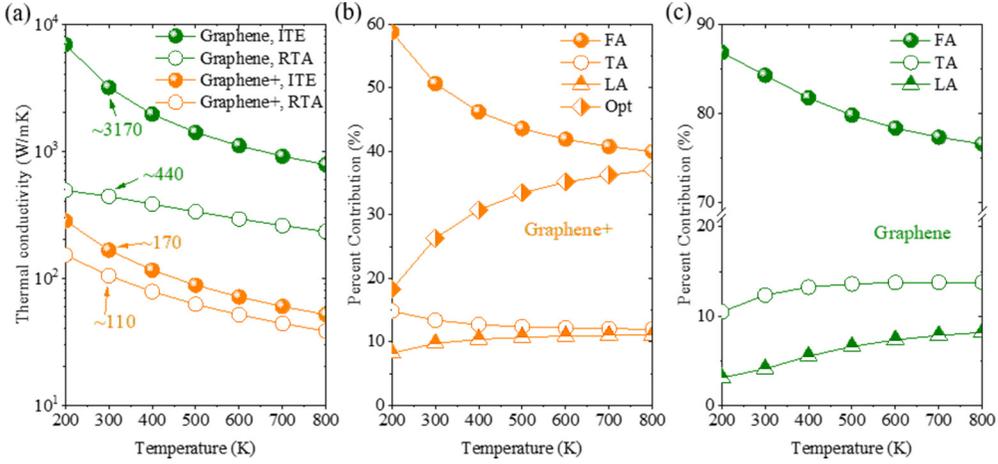

FIG. 3. Thermal transport properties as a function of temperature. (a) Lattice thermal conductivity of graphene+ and graphene as a function of temperature. The percentage contribution of each phonon branch to thermal conductivity as a function of temperature for (b) graphene+ and (c) graphene. The FA, TA, and LA represent the flexible, transverse, and longitudinal acoustic phonon branches, respectively, and "Opt" represents the sum of the optical phonon branches.

Softened phonon modes (weak lattice vibrations) indicate weakened $sp^2$-$sp^3$ bonding, which can be reflected in mechanical properties. The Young's modulus ($E$) describes the physical quantity of a solid material's ability to resist deformation, and its decrease laterally exhibits softening of the bond. The $E$ of graphene+ is calculated to be 120 N/m, which is much lower than that of graphene (320–350 N/m) [31]. Moreover, the contribution of $sp^2$ and $sp^3$ carbon atoms to the lattice vibrations is revealed by the phonon partial density of states ($p$DOS) [Fig. 2(c)]. It is shown that the vibrations of $sp^2$ atoms cover the entire frequency range, while the $sp^3$ carbon atoms mainly contribute to the phonon vibrational frequency range of 20–35 THz. The reason may lie in the fact that $sp^3$ hybridization has relatively weak bonding to soften the phonon mode. Moreover, direct insight can be gained into weak bonding from the distribution of electrons in the energy bands, as shown in Fig. 2(d). Compared to pure $sp^2$ hybridization in graphene [25,47], the occupancy distribution of electrons in the graphene+ orbitals is more inhomogeneous and thus is likely to provide more anharmonicity.

### D. Temperature effect

To further explore the effect of $sp^2$-$sp^3$ bonding on thermal transport, the temperature-dependent $\kappa$ behavior is obtained for graphene+, in comparison with graphene as shown in Fig. 3(a). The $\kappa$ based on the ITE method is ∼170 and ∼3170 W/mK for graphene+ and graphene, respectively. The $\kappa$ of graphene calculated in this work agrees well with the previously reported results of 3000–5000 W/mK from both theoretical and experimental studies [14,24,56,57]. The $\kappa$ of graphene+ is one order of magnitude lower than that of graphene, and also lower than well-known high-$\kappa$ materials such as diamond (∼2000 W/mK) [58], $c$-BAs (∼1000 W/mK) [55,59–61], and TaN (∼900 W/mK) [62,63], but still higher than $MoS_2$ (23.15 W/mK) [64], silicene (∼20 W/mK) [30,65], black phosphorene (∼15 W/mK along the zigzag direction and ∼5 W/mK along the armchair direction) [50], etc. The $\kappa$ calculated by the RTA method is 440 W/mK for graphene, which is almost one order of magnitude underestimated compared to the ITE method, revealing the strong phonon hydrodynamic effect in graphene. In contrast, the RTA method evaluates the $\kappa$ of graphene+ to be ∼110 W/mK, which is 35% lower than the iterative $\kappa$, implying weak phonon hydrodynamics in graphene+ as discussed in Sec. III E.

The percentage contributions of different phonon branches to the total $\kappa$ are shown in Figs. 3(b) and 3(c) for graphene+ and graphene, respectively. For graphene, the $\kappa$ is mainly contributed by the FA branch phonons, which almost exceeds 50% in the temperature range of 200–800 K. Especially at room temperature, the percentage contribution of the FA branch exceeds 80%, which is consistent with previous studies [25,47]. In addition, the contribution of TA to $\kappa$ in graphene increases slightly with temperature, but it is not enough to completely dominate the $\kappa$. As for the optical branch, it con-tributes little to the $\kappa$. Similar to graphene, the FA branch in graphene+ also occupies a slightly larger contribution compared to the other branches. At room temperature, The FA contribution exceeds 50% and is almost decisive for $\kappa$, but it is still weaker than that of graphene. Most differently, the contribution of the optical branch in graphene+ increases significantly with increasing temperature, which contributes almost as much as the FA branch (∼40%) at 800 K. The enhancement of optical phonon branches would empower graphene+ to efficiently dissipate heat at high temperatures [66,67].

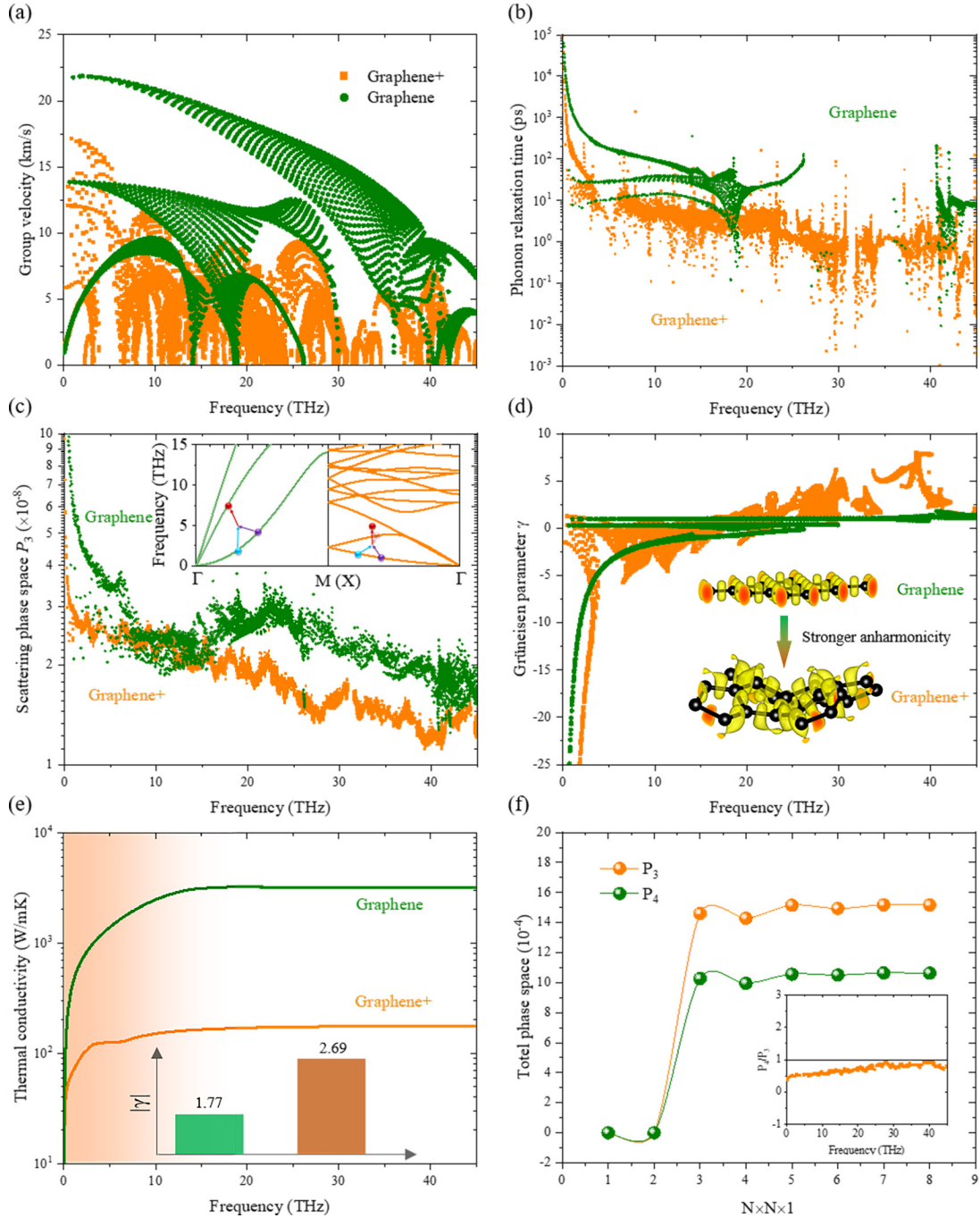

FIG. 4. The mode-level (a) group velocity, (b) relaxation time, (c) scattering phase space, and (d) Grüneisen parameter at 300 K. The insets of (c) and (d) are low-frequency phonon scattering and three-dimensional electron localization functions, respectively. (e) Cumulative thermal conductivity with frequency. The insets are the total Grüneisen parameter. (f) Comparison of the three- and four-phonon scattering phase spaces of graphene+. The inset is the frequency-resolved ratio of the scattering phase space for three- to four-phonon processes in graphene+.

### E. Modal phonon transport properties

To reveal the origin of the large reduction in the $\kappa$ of graphene+ compared to graphene, the phonon properties of different modes over the entire frequency vibrational range are comprehensively investigated. The group velocity is derived from the derivation of the phonon dispersion with respect to the phonon wave number:

$$v = \frac{\partial \omega}{\partial q}. \quad (5)$$

The phonon relaxation time is derived from the reciprocal of scattering rate, i.e., $\Gamma = \frac{1}{\tau}$. Based on Matthiessen's rule [25], summing different scattering events gives the total scattering rate as follows:

$$\frac{1}{\tau(\vec{q}, p)} = \frac{1}{\tau^{\text{anh}}(\vec{q}, p)} + \frac{1}{\tau^{\text{iso}}(\vec{q}, p)} + \frac{1}{\tau^{B}(\vec{q}, p)}, \quad (6)$$

where $1/\tau^{\text{anh}}$, $1/\tau^{\text{iso}}$, and $1/\tau^{B}$ are anharmonic (three-phonon), isotopic, and boundary scattering, respectively.

The results of mode-level phonon properties are plotted in Fig. 4. The group velocity measures the speed of phonons in thermal transport, where higher group velocity generally means higher $\kappa$. As shown in Fig. 4(a), the group velocity of graphene is much higher than that of graphene+. For instance, the highest group velocity of graphene+ is lower than 20 km/s, while graphene is higher. In addition, the group velocity of phonons with frequencies of 0–10 THz in graphene increases rapidly and is generally higher than 5 km/s. However, such a phenomenon is not observed in graphene+, which is caused by the strong softening of the FA branch in graphene+ as previously discussed in Fig. 2(b).

As shown in Fig. 4(b), the relaxation time of graphene is much higher than that of graphene+, which is almost one order of magnitude higher. In particular, the relaxation time of graphene+ tends to decrease sharply in the range of 0–7 THz, which makes the relaxation time of graphene+ nearly two orders of magnitude lower than that of graphene. Namely, strong phonon-phonon scattering exists in the low-frequency region for graphene+. To measure phonon-phonon scattering, the mode-level scattering phase space and the Grüneisen parameters are provided in Figs. 4(c) and 4(d), respectively. Commonly, buckling structures tend to result in large scattering phase spaces [29,68], allowing for a high possibility of scattering between phonons. For instance, a low $\kappa$ of ∼20 W/mK is found in silicene due to the fact that the out-of-plane buckling structure breaks the symmetry [68,69]. In addition, penta-graphene is also found to have a larger phase space, which contributes to the low $\kappa$ [56]. In contrast, Fig. 4(c) reveals that graphene+ possesses a smaller scattering phase space compared with graphene, and its origin can be traced back to its unique phonon beam effect, as shown in the inset of Fig. 4(c). Since the phonon branches are close to each other, this suppresses the direct scattering between phonons on adjacent branches due to not satisfying energy conservation. Considering the significant difference in atomic number, we provide a comparison of scattering phase space tests with the same atomic number (see Fig. S2 in the Supplemental Material [54]), where a larger scattering phase space is still observed in graphene. The results are consistent with the analysis above. Thus, the strong phonon-phonon scattering of graphene+ must originate from the large Grüneisen parameter representing strong anharmonicity considering the small scattering phase space. Larger mode-level [Fig. 4(d)] and total [the inset in Fig. 4(e)] Grüneisen parameters imply high scattering intensity, revealing strong anharmonicity. The strong anharmonicity is attributed to the softened phonon modes, which overcomes the small scattering phase space and leads to the large phonon scattering rate.

The differences in the phonon properties can be fundamentally analyzed from the mode-level $\kappa$ as shown in Fig. 4(e). The cumulative $\kappa$ of graphene and graphene+ increases rapidly with increasing frequency. When the phonon frequency lies in the range 0–10 THz, the $\kappa$ of graphene increases the fastest, indicating phonons with frequencies of 0–10 THz dominate the lattice vibrations. Hence, the reduced $\kappa$ of graphene+ lies in the softened low-frequency acoustic phonons, which is consistent with the contribution from phonon branches (Fig. 3). Note that the dominant frequency range of the $\kappa$ contribution is the same as the vibrational range of $sp^2$ atoms [Fig. 2(c)]. Therefore, the thermal transport is mainly governed by the phonons from the $sp^2$ carbon atomic vibrations in graphene+ while the existence of $sp^3$ carbon atoms will suppress the $\kappa$.

The discussion above treats three-phonon scattering as the dominant scattering process. However, four-phonon scattering may play an important role. Typically, strong four-phonon scattering mainly originates from the phonon bands decoupled by the wide phonon band gap [59–61,70]. For instance, the wide phonon band gap in $c$-BAs leads to a much larger phase space for four-phonon scattering ($P_4 = 3.9 \times 10^{-3}$) than three-phonon scattering ($P_3 = 3.8 \times 10^{-4}$). To evaluate the four-phonon effect in graphene+, the comparison of the $P_4$ and $P_3$ as a function of the $N \times N \times 1$ grid is provided in Fig. 4(f). It is shown in graphene+ that the $P_3$ is significantly higher than $P_4$, which means that three-phonon scattering dominates the phonon-phonon scattering process. In addition, the mode-level scattering phase space ratio $P_4/P_3$ is calculated to measure the effect of four-phonon scattering in graphene+ as shown in the inset of Fig. 4(f). The ratio $P_4/P_3$ below 1 exhibits a weak four-phonon process, especially at low frequencies, in sharp contrast to $c$-BAs ($P_4/P_3 \sim 10$). The weaker four-phonon effect in graphene+ is due to the narrow phonon band gap from the same atomic mass, while the wide band gap is induced by the large atomic mass difference between As and B atoms in $c$-BAs [55]. Therefore, three-phonon scattering dominates thermal transport for all the phonons in graphene+, and the $\kappa$ calculated based on three-phonon scattering is acceptable.

### F. Phonon hydrodynamic effect

Phonon hydrodynamics [71,72] in graphene are closely related to size effects, which can also be captured in transition-metal chalcogenides [73]. Here, the cumulative $\kappa$ as a function of mean free path (MFP) is plotted in Fig. 5(a). Based on the phonon gas model, the $\kappa$ can be obtained as $\kappa \sim \frac{1}{3}Cvl$, where $C$, $v$, and $l$ represent the specific-heat capacity, group velocity, and MFP, respectively. Phonons with larger MFP tend to be less scattered, which contributes to higher $\kappa$. The MFP of graphene+ is in the range of $10-10^5$ nm, slightly lower than graphene ($10^2-10^5$ nm). Further, the MFP corresponding to 50% of the cumulative $\kappa$ is selected as the representative MFP (rMFP) to characterize the size effect on thermal transport.

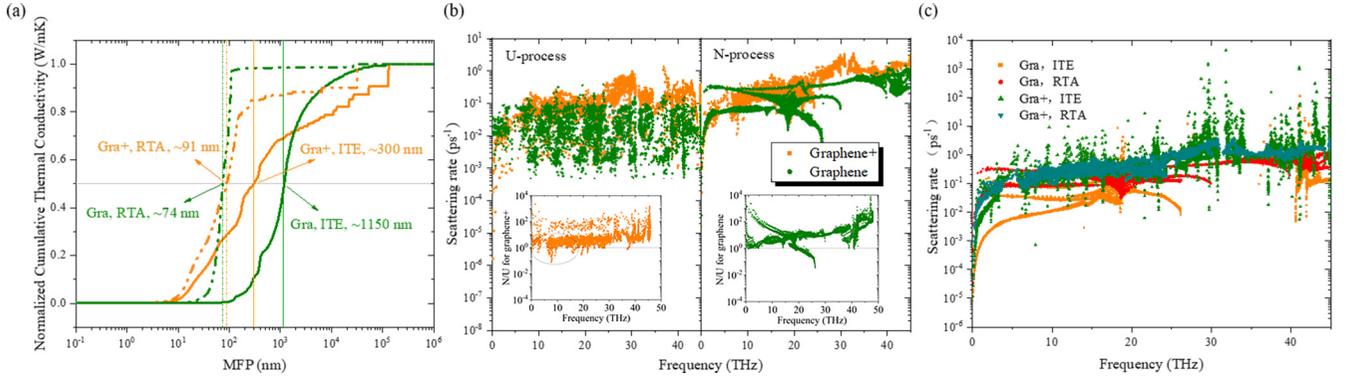

FIG. 5. (a) Normalized cumulative thermal conductivity at 300 K as a function of mean free path (MFP). (b) Comparative analysis of $N$ and $U$ processes in graphene+ and graphene. The inset is the ratio of the scattering rates of $N$ to $U$ processes (N/U). Orange and green represent graphene+ and graphene, respectively. (c) Comparison of phonon scattering rates based on iterative (ITE) and relaxation time approximation (RTA) methods for graphene+ (Gra+) and graphene (Gra).

The rMFP of graphene is ∼1150 nm, consistent with previous reports [25]. Due to the smaller group velocity and relaxation time, the rMFP of ∼300 nm for graphene+ is one order of magnitude lower than graphene. This means that the softened bonding results in smaller rMFPs and narrower MFP distribution with frequency, which leads to a shift in the motion pattern of phonons during transport, as will be discussed below.

Phonon hydrodynamics means that phonons exhibit fluidlike macroscopic drift motions, namely the momentum-conserving collisions between a large number of phonons, which leads to the hydrodynamic flow for thermal transport. In such collisions, a freezing phonon reduces the wavelength of thermally excited phonons, thereby preserving momentum, leading to normal ($N$) scattering events. On the contrary, when the wave vector generated by scattering between two phonons exceeds the unit vector of the reciprocal lattice, the excess momentum is lost to the underlying lattice. Such phenomena are defined as umklapp ($U$) scattering events, because they require sufficiently large wave vectors. The phonon hydrodynamics can be evaluated by the collision of the $N$ and $U$ processes, where the phonon scattering process is determined by the conservation of momentum [47]:

$$\vec{q} \pm \vec{q'} = \vec{q''} + \vec{K}, \quad (7)$$

where $\vec{K} = 0$ and $\vec{K} \neq 0$ correspond to $N$ and $U$ processes, respectively.

The quadratic FA phonon branch in graphene leads to the large scattering phase space of the $N$ process overtaking the $U$ process, which contributes to the strong hydrodynamic effect [57]. In graphene+, the softened $sp^2$-$sp^3$ bond suppresses the acoustic phonon branches, making them flatter, which leads to a predictably weak hydrodynamic effect. The phonon-phonon scattering rates contributed by $N$ and $U$ are plotted in Fig. 5(b). In both graphene and graphene+, the $N$ process contributes to a stronger scattering rate than the $U$ process. Note that the contribution of the $N$ process to $\kappa$ in graphene+ is weaker than that in graphene due to the strong $N$ process scattering. The relative strength of $N$ and $U$ processes can be measured by their mode-level ratio (N/U) as shown in the inset of Fig. 5(c). Obviously, graphene+ exhibits a weaker phonon hydrodynamic effect compared to graphene, because the N/U ratio is larger than 1 in the region of 0–10 THz where phonons dominate the thermal transport. The results are consistent with the phenomena as revealed in Fig. 1(f) that the introduction of $sp^3$ hybridization significantly suppresses the phonon hydrodynamic effect compared to $sp^2$ hybridization.

Another intuitive evidence of hydrodynamics is the phonon Poiseuille flow, which can be quantified by the exponential power of $\kappa$ as a function of temperature [74,75]. By fitting the $\kappa$ versus temperature relationship $\kappa \sim T^{-c}$, $c = 1.3$ is obtained for graphene+ but $c = 1.8$ is for graphene. A small exponential factor means weaker Poiseuille flow, i.e., weak hydrodynamics. In graphene, long-range phonons ensure that phonon scattering is achieved in reciprocal space ($N$ process). On contrary, the reduced MFP tends to make phonons more inclined to complete $U$ process scattering across the Brillouin zone in graphene+. As shown in Fig. 5(a), the rMFP obtained by the ITE method in graphene (RTA: ∼74 → ITE: ∼1150 nm) is two orders of magnitude larger than the RTA method, while the rMFP of graphene+ is one order of magnitude larger (RTA: ∼91 → ITE: ∼300 nm). The strong hydrodynamics leads to a significant improvement in rMFP for graphene: $\Delta \text{MFP}_{\text{gra}} = 14.5 > \Delta \text{MFP}_{\text{gra+}} = 2.3$, where $\Delta \text{MFP} = \frac{\text{rMFP}_{\text{ITE}} - \text{rMFP}_{\text{RTA}}}{\text{rMFP}_{\text{RTA}}}$. Because the RTA method treats the $N$-process as thermal resistance, RTA-based phonon-phonon scattering is severely overestimated in graphene, as shown in Fig. 5(c), which leads to a significant underestimation of $\kappa$.

### G. Quantification of interatomic bonding

Although the above analysis explains that weak bonding causes strong anharmonicity and weak hydrodynamic effects, the weakened $sp^2$-$sp^3$ network bonding is not well quantified. The restoring force of lattice atoms when vibrating near the equilibrium position can be evaluated by the potential well, which reflects the strength of the bonding between atoms [25,76]. The potential well describes the relationship between the potential energy of the system and the positive and negative atomic displacements. As shown in Fig. 6(a), at the same potential level, the wider potential well of graphene+ means that it needs more energy to recover to the equilibrium position, which indicates the small restoring force and the weak atomic bonding.

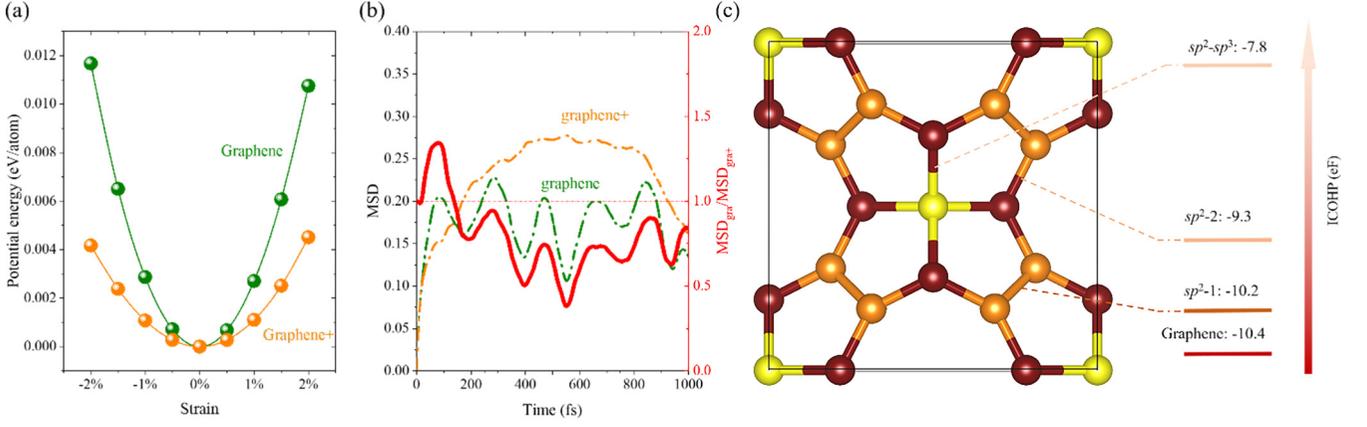

FIG. 6. (a) Comparison of potential wells. (b) Root-mean-square displacement (MSD) calculated from the *ab initio* molecular-dynamics simulations. (c) The integrated crystal orbital Hamilton population (ICOHP) for different hybrid bonds.

The weak restoring force and bond strength can be further evaluated by the mean-square displacement (MSD), which describes the amplitude of the lattice vibration at the equilibrium position and its sensitivity to the potential energy surface. MSD can be obtained by tracking the evolution of atomic positions in the system with time based on molecular-dynamics simulations, which is derived by [77]

$$MSD(m) = \frac{1}{N_{\text{particles}}} \sum_{i=1}^{N_{\text{particles}}} \frac{1}{N-m}$$
$$\times \sum_{k=0}^{N-m-1} (\mathbf{r}_i(k+m) - \mathbf{r}_i(k))^2, \quad (8)$$

where $N_{\text{particles}}$, $N$, and $\mathbf{r}_i(t)$ are the number of atoms in the system, the number of frames in the simulation, and the position of atom $i$ at simulation time $t$, respectively. Considering the poor precision of the empirical potential, *ab initio* molecular-dynamics (AIMD) simulations are performed to obtain the MSDs of graphene+ and graphene as comparably shown in Fig. 6(b). The MSD ratio ($\text{MSD}_{\text{gra}}/\text{MSD}_{\text{gra+}}$) of graphene ($\text{MSD}_{\text{gra}}$) and graphene+ ($\text{MSD}_{\text{gra+}}$) is also plotted to assess the relative magnitudes within the simulation domain. Near the equilibrium position, the energy of atomic vibrations originates from thermal fluctuations ($k_B T$), leading to a deviation. In the simulations, a larger MSD than graphene is found in graphene+, implying that it requires more energy to recover further deviations due to a wider potential well.

Based on the above discussions, weaker bonding is identified in graphene+ compared to graphene. The bonding form in graphene+ is more complex, where both $sp^2$- and $sp^3$-hybridized carbon atoms exist, while only $sp^2$-hybridized carbon atoms exist in graphene. Hence, the final step is to determine the relative strength of the bonds between the different hybridized atoms. As shown in Fig. 6(c), the bonding strength between different hybridized atoms in graphene+ is quantified by the integrated crystal orbital Hamilton population (ICOHP) [78,79]. Smaller ICOHP indicates larger bonding state occupation by orbital electrons, i.e., stronger interatomic bonding. For graphene, the ICOHP of pure $sp^2$ bonds is $-10.4$ eF. As for graphene+, it possesses three types of bonding, including $sp^2$-$sp^3$, $sp^2$-1, and $sp^2$-2 bonds. Notably, the ICOHP of an $sp^2$-$sp^3$ ($-7.8$ eF) bond is significantly higher than an $sp^2$ bond in graphene, while $sp^2$-1 ($-10.2$ eF) and $sp^2$-2 ($-9.3$ eF) bonds hardly change. Hence, the origin of the one order of magnitude lower $\kappa$ than graphene in graphene+ can be fundamentally explained by the softened $sp^2$-$sp^3$ bonding. In graphene allotropes with $sp^2$-hybridized atoms, the change in $\kappa$ is independent of the bonding strength, but it depends on the band folding induced by the unique phononic crystal structure [26], which cannot explain the decrease in $\kappa$ of graphene+ due to $sp^3$ hybridization. In graphene+, the $sp^3$ hybridization softens the bonds, leading to phonon-phonon scattering, especially those adjacent to $sp^3$-hybridized carbon atoms ($sp^2$-$sp^3$ bond).

## IV. CONCLUSION

In summary, we have investigated the thermal transport properties of the novel carbon-based monolayer graphene+ with $sp^2$-$sp^3$ hybridization from the state-of-the-art first-principles. Our work reveals the effect of the introduction of $sp^3$ hybridization on phonon thermal transport, which leads to softened $sp^2$-$sp^3$ bonding while $sp^2$ bonding strength hardly changes in graphene allotropes. It is found that the softened $sp^2$-$sp^3$ bonding suppresses the low-frequency phonons (0–10 THz) that dominate $\kappa$ of graphene+. Further, the softened phonon modes induce strong anharmonicity and weak phonon hydrodynamic effects. Based on the analysis of mode-level phonon properties, the strong anharmonicity in graphene+ can be attributed to the large Grüneisen parameter, which overcomes the small scattering phase space and thus leads to the low relaxation time. It can be revealed intuitively through the inhomogeneous electron distribution in the band structures and ELF. Weak hydrodynamic effects are captured by the phonon collisions based on the momentum conservation process ($N$ and $U$ process) and the exponential factor $c$ of the phonon Poiseuille flow. Smaller phonon MFPs ensure a weaker $N$ process, and a smaller $c$ value indicates weaker phonon Poiseuille flow. Hence, stronger phonon-phonon scattering is found in graphene+ due to weaker hydrodynamics. Finally, we quantify the strength of interatomic bonding based on potential wells and MSD, and we identify the strength of $sp^2$-$sp^3$ and $sp^2$ bonding by ICOHP. The $sp^3$ atoms would severely soften the adjacent $sp^2$-$sp^3$ bonding and contribute to the low $\kappa$.

The comprehensive analysis of the factors (see Table S1 in the Supplemental Material [54]) describing changes in thermal transport provides clear physical insights into the effect of hybridized $sp^2$-$sp^3$ bonding in graphene+ on thermal transport behavior. Due to small scattering phase space like graphene, graphene+ is expected to serve as an ideal model to study atomic bonding versus thermal transport properties. Combined with the unique mechanical auxetic and electronic Dirac properties, an in-depth understanding of the thermal transport behavior for graphene+ can provide guidance for electronic devices and related thermal management applications.

## ACKNOWLEDGMENTS

This work is supported by the National Natural Science Foundation of China (Grants No. 52006057, No. 51906097, and No. 11904324), the Fundamental Research Funds for the Central Universities (Grants No. 531119200237 and No. 541109010001), and the State Key Laboratory of Advanced Design and Manufacturing for Vehicle Body at Hunan University (Grant No. 52175013). The numerical calculations in this paper have been done on the supercomputing system of the National Supercomputing Center in Changsha, and the Hefei advanced computing center. The authors would like to thank Rongkun Chen (Kunming University of Science and Technology) for helpful discussions.

G.Q. supervised the project. L.Y. performed all the calculations and analysis. All the authors contributed to interpreting the results. The paper was written by L.Y., with contributions from all the authors.

The authors declare no competing interests.

[1] R. Hoffmann, A. A. Kabanov, A. A. Golov, and D. M. Proserpio, Homo citans and carbon allotropes: For an ethics of citation, Angew. Chem. Int. Ed. **55**, 10962 (2016).

[2] E. A. Ekimov, V. A. Sidorov, E. D. Bauer, N. N. Mel'nik, N. J. Curro, J. D. Thompson, and S. M. Stishov, Superconductivity in diamond, Nature (London) **428**, 6982 (2004).

[3] X.-L. Sheng, Q.-B. Yan, F. Ye, Q.-R. Zheng, and G. Su, T-Carbon: A Novel Carbon Allotrope, Phys. Rev. Lett. **106**, 155703 (2011).

[4] Y. Xu, M. Hu, Z. Wang, J. Liang, J. Li, X. Zhu, and M. Wang, A new metallic porous carbon phase TP-C12 with an $sp^2$-$sp^3$ bonding network: A first-principle calculation, ChemistrySelect **3**, 8402 (2018).

[5] D. Malko, C. Neiss, F. Viñes, and A. Görling, Competition for Graphene: Graphynes with Direction-Dependent Dirac Cones, Phys. Rev. Lett. **108**, 086804 (2012).

[6] H. Yin, X. Shi, C. He, M. Martinez-Canales, J. Li, C. J. Pickard, C. Tang, T. Ouyang, C. Zhang, and J. Zhong, Stone-Wales graphene: A two-dimensional carbon semimetal with magic stability, Phys. Rev. B **99**, 041405 (2019).

[7] M. Takagi, T. Taketsugu, H. Kino, Y. Tateyama, K. Terakura, and S. Maeda, Global search for low-lying crystal structures using the artificial force induced reaction method: A case study on carbon, Phys. Rev. B **95**, 184110 (2017).

[8] M. Xu, T. Liang, M. Shi, and H. Chen, Graphene-like two-dimensional materials, Chem. Rev. **113**, 3766 (2013).

[9] C. He, X. Shi, S. J. Clark, J. Li, C. J. Pickard, T. Ouyang, C. Zhang, C. Tang, and J. Zhong, Complex Low Energy Tetrahedral Polymorphs of Group IV Elements from First Principles, Phys. Rev. Lett. **121**, 175701 (2018).

[10] X. Yang, M. Yao, X. Wu, S. Liu, S. Chen, K. Yang, R. Liu, T. Cui, B. Sundqvist, and B. Liu, Novel Superhard $sp^3$ Carbon Allotrope from Cold-compressed $C_{70}$ Peapods, Phys. Rev. Lett. **118**, 245701 (2017).

[11] Z. Zhao, B. Xu, X.-F. Zhou, L.-M. Wang, B. Wen, J. He, Z. Liu, H.-T. Wang, and Y. Tian, Novel Superhard Carbon: C-centered Orthorhombic $C_8$, Phys. Rev. Lett. **107**, 215502 (2011).

[12] K. S. Novoselov, A. K. Geim, S. V. Morozov, D. Jiang, M. I. Katsnelson, I. V. Grigorieva, S. V. Dubonos, and A. A. Firsov, Two-dimensional gas of massless Dirac fermions in graphene, Nature (London) **438**, 7065 (2005).

[13] C. Lee, X. Wei, J. W. Kysar, and J. Hone, Measurement of the elastic properties and intrinsic strength of monolayer graphene, Science **321**, 385 (2008).

[14] S. Chen, Q. Wu, C. Mishra, J. Kang, H. Zhang, K. Cho, W. Cai, A. A. Balandin, and R. S. Ruoff, Thermal conductivity of isotopically modified graphene, Nat. Mater **11**, 3 (2012).

[15] S. Zhang, J. Zhou, Q. Wang, X. Chen, Y. Kawazoe, and P. Jena, Penta-Graphene: A new carbon allotrope, Proc. Natl. Acad. Sci. USA **112**, 2372 (2015).

[16] S. Sun, F. Meng, Y. Xu, J. He, Y. Ni, and H. Wang, Flexible, auxetic and strain-tunable two dimensional Penta-$X_2$C family as water splitting photocatalysts with high carrier mobility, J. Mater. Chem. A **7**, 7791 (2019).

[17] G. Liu, Q. Zeng, P. Zhu, R. Quhe, and P. Lu, Negative poisson's ratio in monolayer $PdSe_2$, Comput. Mater. Sci. **160**, 309 (2019).

[18] X. Lv, L. Yu, F. Li, J. Gong, Y. He, and Z. Chen, Penta-$MS_2$ (M = Mn, Ni, Cu/Ag and Zn/Cd) monolayers with negative poisson's ratios and tunable bandgaps as water-splitting photocatalysts, J. Mater. Chem. A **9**, 6993 (2021).

[19] W.-L. Tao, Y.-Q. Zhao, Z.-Y. Zeng, X.-R. Chen, and H.-Y. Geng, Anisotropic thermoelectric materials: Pentagonal $PtM_2$ (M = S, Se, Te), ACS Appl. Mater. Interfaces **13**, 8700 (2021).

[20] Y. Liu, G. Wang, Q. Huang, L. Guo, and X. Chen, Structural and Electronic Properties of *T* Graphene: A Two-Dimensional Carbon Allotrope with Tetrarings, Phys. Rev. Lett. **108**, 225505 (2012).

[21] J.-W. Jiang, J. Leng, J. Li, Z. Guo, T. Chang, X. Guo, and T. Zhang, Twin graphene: A novel two-dimensional semiconducting carbon allotrope, Carbon **118**, 370 (2017).

[22] H. P. Veeravenkata and A. Jain, Density functional theory driven phononic thermal conductivity prediction of biphenylene: A comparison with graphene, Carbon **183**, 893 (2021).

[23] Q. Fan, L. Yan, M. Tripp, O. Krejcí, S. Dimosthenous, S. Kachel, M. Chen, A. Foster, U. Koert, P. Liljeroth *et al.*, Biphenylene network: A nonbenzenoid carbon allotrope, Science **372**, 852 (2021).

[24] B. Peng, D. Zhang, H. Zhang, H. Shao, G. Ni, Y. Zhu, and H. Zhu, The conflicting role of buckled structure in phonon transport of 2D group-IV and group-V materials, Nanoscale **9**, 7397 (2017).

[25] Z. Qin, G. Qin, X. Zuo, Z. Xiong, and M. Hu, Orbitally driven low thermal conductivity of monolayer gallium nitride (GaN) with planar honeycomb Structure: A comparative study, Nanoscale **9**, 4295 (2017).

[26] U. Choudhry, S. Yue, and B. Liao, Origins of significant reduction of lattice thermal conductivity in graphene allotropes, Phys. Rev. B **100**, 165401 (2019).

[27] F. Q. Wang, J. Yu, Q. Wang, Y. Kawazoe, and P. Jena, Lattice thermal conductivity of penta-graphene, Carbon **105**, 424 (2016).

[28] L. Yu, Y. Zhan, D. Wei, C. Shen, H. Zhang, Z. Qin, and G. Qin, Multifunctional two-dimensional graphene-like boron nitride allotrope of g-$B_3N_5$: A competitor to g-BN? J. Alloys Compd. **921**, 165913 (2022).

[29] L. Yu, Y. Tian, X. Zheng, H. Wang, C. Shen, and G. Qin, Abnormal enhancement of thermal conductivity by planar structure: A comparative study of graphene-like materials, Int. J. Therm. Sci. **174**, 107438 (2022).

[30] B. Peng, H. Zhang, H. Shao, Y. Xu, G. Ni, R. Zhang, and H. Zhu, Phonon transport properties of two-dimensional group-IV materials from Ab initio calculations, Phys. Rev. B **94**, 245420 (2016).

[31] L. Yu, Z. Qin, H. Wang, X. Zheng, and G. Qin, Half-negative poisson's ratio in graphene+ with intrinsic dirac nodal loop, Cell Rep. Phys. Sci. **3**, 100790 (2022).

[32] S. Wang, Y. Si, B. Yang, E. Ruckenstein, and H. Chen, Two-dimensional carbon-based auxetic materials for broad-spectrum metal-ion battery anodes, J. Phys. Chem. Lett. **10**, 3269 (2019).

[33] R. Peng, Y. Ma, Q. Wu, B. Huang, and Y. Dai, Two-dimensional materials with intrinsic auxeticity: Progress and perspectives, Nanoscale **11**, 11413 (2019).

[34] Z. Wang and H. Hu, Auxetic materials and their potential applications in textiles, Text. Res. J. **84**, 1600 (2014).

[35] H. Pan, Z. Li, C.-C. Liu, G. Zhu, Z. Qiao, and Y. Yao, Valley-Polarized Quantum Anomalous Hall Effect in Silicene, Phys. Rev. Lett. **112**, 106802 (2014).

[36] W.-K. Tse, Z. Qiao, Y. Yao, A. H. MacDonald, and Q. Niu, Quantum anomalous Hall effect in single-layer and bilayer graphene, Phys. Rev. B **83**, 155447 (2011).

[37] L. Banszerus, M. Schmitz, S. Engels, J. Dauber, M. Oellers, F. Haupt, K. Watanabe, T. Taniguchi, B. Beschoten, and C. Stampfer, Ultrahigh-mobility graphene devices from chemical vapor deposition on reusable copper, Sci. Adv. **1**, e1500222 (2015).

[38] L. A. Ponomarenko, R. Yang, T. M. Mohiuddin, M. I. Katsnelson, K. S. Novoselov, S. V. Morozov, A. A. Zhukov, F. Schedin, E. W. Hill, and A. K. Geim, Effect of a High-$\kappa$ Environment on Charge Carrier Mobility in Graphene, Phys. Rev. Lett. **102**, 206603 (2009).

[39] C.-T. Toh, H. Zhang, J. Lin, A. Mayorov, Y. Wang, C. Orofeo, D. B. Ferry, H. Andersen, N. Kakenov, Z. Guo *et al.*, Synthesis and properties of free-standing monolayer amorphous carbon, Nature (London) **577**, 7789 (2020).

[40] L. Hou, X. Cui, B. Guan, S. Wang, R. Li, Y. Liu, D. Zhu, and J. Zheng, Synthesis of a monolayer fullerene network, Nature (London) **606**, 7914 (2022).

[41] J. Zhang, R. X. Zhu, A. pan, C. Han, X. li, D. Zhao, C. MA, W. Wang, H. Su *et al.*, Pseudo-topotactic conversion of carbon nanotubes to t-carbon nanowires under picosecond laser irradiation in methanol, Nat. Commun. **8**, 1 (2017).

[42] G. Kresse and J. Furthmüller, Efficient iterative schemes for ab initio total-energy calculations using a plane-wave basis set, Phys. Rev. B **54**, 11169 (1996).

[43] G. Kresse and J. Hafner, *Ab initio* molecular-dynamics simulation of the liquid-metal–amorphous-semiconductor transition in germanium, Phys. Rev. B **49**, 14251 (1994).

[44] J. P. Perdew, K. Burke, and M. Ernzerhof, Generalized Gradient Approximation Made Simple, Phys. Rev. Lett. **77**, 3865 (1996).

[45] H. J. Monkhorst and J. D. Pack, Special points for Brillouin-zone integrations, Phys. Rev. B **13**, 5188 (1976).

[46] W. Li, J. Carrete, N. A. Katcho, and N. Mingo, ShengBTE: A solver of the boltzmann transport equation for phonons, Comput. Phys. Commun. **185**, 1747 (2014).

[47] H. Wang, L. Yu, J. Xu, D. Wei, G. Qin, Y. Yao, and M. Hu, Intrinsically low lattice thermal conductivity of monolayer hexagonal aluminum nitride (h-AlN) from first-principles: A comparative study with graphene, Int. J. Therm. Sci. **162**, 106772 (2021).

[48] S. Lee, K. Esfarjani, T. Luo, J. Zhou, Z. Tian, and G. Chen, Resonant bonding leads to low lattice thermal conductivity, Nat. Commun. **5**, 3525 (2014).

[49] G. Qin and M. Hu, Accelerating evaluation of converged lattice thermal conductivity, npj Comput. Mater. **4**, 3 (2018).

[50] G. Qin, X. Zhang, S.-Y. Yue, Z. Qin, H. Wang, Y. Han, and M. Hu, Resonant bonding driven giant phonon anharmonicity and low thermal conductivity of phosphorene, Phys. Rev. B **94**, 165445 (2016).

[51] B.-T. Wang, P.-F. Liu, J.-J. Zheng, W. Yin, and F. Wang, First-principles study of superconductivity in the two- and three-dimensional forms of $PbTiSe_2$: Suppressed charge density wave in $1T-TiSe_2$, Phys. Rev. B **98**, 014514 (2018).

[52] O. Delaire, J. Ma, K. Marty, A. F. May, M. A. McGuire, M-H. Du, D. J. Singh, A. Podlesnyak, G. Ehlers, M. D. Lumsden *et al.*, Giant anharmonic phonon scattering in PbTe, Nat. Mater **10**, 614 (2011).

[53] K. Shportko, S. Kremers, M. Woda, D. Lencer, J. Robertson, and M. Wuttig, Resonant bonding in crystalline phase-change materials, Nat. Mater **7**, 653 (2008).

[54] See Supplemental Material at http://link.aps.org/supplemental/10.1103/PhysRevB.106.125410 for Note S1 for energy conservation, Table S1 for the effect of different factors, Fig. S1 for phonon dispersions, Fig. S2 for scattering phase space, and Fig. S3 for scattering channels.

[55] L. Lindsay, D. A. Broido, and T. L. Reinecke, First-Principles Determination of Ultrahigh Thermal Conductivity of Boron Arsenide: A Competitor for Diamond? Phys. Rev. Lett. **111**, 025901 (2013).

[56] X. Wu, V. Varshney, J. Lee, T. Zhang, J. L. Wohlwend, A. K. Roy, and T. Luo, Hydrogenation of penta-graphene leads to unexpected large improvement in thermal conductivity, Nano Lett. **16**, 3925 (2016).

[57] E. Muñoz, J. Lu, and B. I. Yakobson, Ballistic thermal conductance of graphene ribbons, Nano Lett. **10**, 1652 (2010).

[58] G. A. Slack, Thermal conductivity of pure and impure silicon, silicon carbide, and diamond, J. Appl. Phys. **35**, 3460 (1964).

[59] S. Li, Q. Zheng, Y. Lv, X. Liu, X. Wang, P. Y. Huang, D. G. Cahill, and B. Lv, High thermal conductivity in cubic boron arsenide crystals, Science **361**, 579 (2018).

[60] J. S. Kang, M. Li, H. Wu, H. Nguyen, and Y. Hu, Experimental observation of high thermal conductivity in boron arsenide, Science **361**, 575 (2018).

[61] F. Tian, B. Song, X. Chen, N. K. Ravichandran, Y. Lv, K. Chen, S. Sullivan, J. Kim, Y. Zhou, T. Liu *et al.*, Unusual high thermal conductivity in boron arsenide bulk crystals, Science **361**, 582 (2018).

[62] A. Kundu, X. Yang, J. Ma, T. Feng, J. Carrete, X. Ruan, G. K. H. Madsen, and W. Li, Ultrahigh Thermal Conductivity of Θ-Phase Tantalum Nitride, Phys. Rev. Lett. **126**, 115901 (2021).

[63] S.-D. Guo and B.-G. Liu, Ultrahigh lattice thermal conductivity in topological semimetal TaN caused by a large acoustic-optical gap, J. Phys.: Condens. Matter **30**, 105701 (2018).

[64] S.-D. Guo, Phonon transport in janus monolayer MoSSe: A first-principles study, Phys. Chem. Chem. Phys. **20**, 7236 (2018).

[65] A. Taheri, C. Da Silva, and C. H. Amon, Phonon thermal transport in $\beta$-N x (X = P, As, Sb) monolayers: A first-principles study of the interplay between harmonic and anharmonic phonon properties, Phys. Rev. B **99**, 235425 (2019).

[66] G. Qin, Z. Qin, H. Wang, and M. Hu, Anomalously temperature-dependent thermal conductivity of monolayer GaN with large deviations from the traditional $1/T$ law, Phys. Rev. B **95**, 195416 (2017).

[67] H. Wang, G. Qin, G. Li, Q. Wang, and M. Hu, Low thermal conductivity of monolayer ZnO and its anomalous temperature dependence, Phys. Chem. Chem. Phys. **19**, 12882 (2017).

[68] X. Gu and R. Yang, First-principles prediction of phononic thermal conductivity of silicene: A comparison with graphene, J. Appl. Phys. **117**, 025102 (2015).

[69] H. Xie, M. Hu, and H. Bao, Thermal conductivity of silicene from first-principles, Appl. Phys. Lett. **104**, 131906(R) (2014).

[70] T. Feng, L. Lindsay, and X. Ruan, Four-phonon scattering significantly reduces intrinsic thermal conductivity of solids, Phys. Rev. B **96**, 161201 (2017).

[71] S. Huberman, R. A. Duncan, K. Chen, B. Song, V. Chiloyan, Z. Ding, A. A. Maznev, G. Chen, and K. A. Nelson, Observation of second sound in graphite at temperatures above 100 k, Science **364**, 375 (2019).

[72] S. Lee, D. Broido, K. Esfarjani, and G. Chen, Hydrodynamic phonon transport in suspended graphene, Nat. Commun. **6**, 6290 (2015).

[73] P. Torres, F. X. Alvarez, X. Cartoixà, and R. Rurali, Thermal conductivity and phonon hydrodynamics in transition metal dichalcogenides from first-principles, 2D Mater. **6**, 035002 (2019).

[74] F. Duan, C. Shen, H. Zhang, and G. Qin, Hydrodynamically Enhanced Thermal Transport Due to Strong Interlayer Interactions: A Case Study of Strained Bilayer Graphene, Phys. Rev. B **105**, 125406 (2022).

[75] R. A. Guyer and J. A. Krumhansl, Thermal conductivity, second sound, and phonon hydrodynamic phenomena in nonmetallic crystals, Phys. Rev. **148**, 778 (1966).

[76] A. Lou, Q.-B. Liu, and H.-H. Fu, Enhanced thermoelectric performance by lone-pair electrons and bond anharmonicity in the two-dimensional $Ge_2Y_2$ family of materials with $y =$ N, P, As, or Sb, Phys. Rev. B **105**, 075431 (2022).

[77] V. Wang, N. Xu, J.-C. Liu, G. Tang, and W.-T. Geng, VASPKIT: A user-friendly interface facilitating high-throughput computing and analysis using VASP code, Comput. Phys. Commun. **267**, 108033 (2021).

[78] R. Dronskowski and P. E. Bloechl, Crystal orbital hamilton populations (COHP): Energy-resolved visualization of chemical bonding in solids based on density-functional calculations, J. Phys. Chem. **97**, 8617 (1993).

[79] V. L. Deringer, A. L. Tchougréeff, and R. Dronskowski, Crystal orbital hamilton population (COHP) analysis as projected from plane-wave basis sets, J. Phys. Chem. A **115**, 5461 (2011).